\documentclass{ws-mpla}

\begin{document}

\title{The Unified Equation of State for Dark Matter and Dark Energy}
\author{WEI WANG, {\footnotesize YUANXING GUI\thanks{
Corresponding author},} \\
SUHONG ZHANG, GUANGHAI GUO, YING SHAO}
\address{Department of Physics, Dalian University of Technology,\\
Dalian 116024, P. R. China\\
thphys@dlut.edu.cn} \maketitle

\begin{abstract}
We assume that dark matter and dark energy satisfy the unified
equation of state: $p=B(z)\rho$, with $p=p_{dE}$,
$\rho=\rho_{dm}+\rho_{dE}$, where the pressure of dark matter
$p_{dm}=0$ has been taken into account. A special function
$B=-\frac{A}{(1+z)^{\alpha }}$ is presented, which can well
describe the evolution of the universe. In this model, the
universe will end up with a Big Rip. By further simple analysis,
we know other choices of the function $B$ can also describe the
universe but lead to a different doomsday.
\end{abstract}


\markboth{Wei Wang and Yuanxing Gui} {The Unified Equation of
State for Dark Matter and Dark Energy}

\address{Department of Physics, Dalian University of Technology\\
Dalian 116024, P.R.China\\
guiyx@dlut.edu.cn}

\ccode{PACS Nos.: 98.80.-k}

\section{Introduction}

The astronomical observations indicate that the expansion of our
universe accelerates\cite{1}. This accelerated expansion is
ascribed to a mysterious component with the negative pressure
called dark energy that comprises about $\frac{2}{3}$ of the
energy density of the universe. The simplest model for the dark
energy is a cosmological constant. There are many more complex
models proposed attempting to explain dark energy either as a
dynamic substance(for example Quintessence\cite{2},
Phantom\cite{3}) or via some forms of modified gravitational
theory perhaps related to extra dimensions and string
physics\cite{4}. Now the unified models for dark energy and dark
matter have been also discussed\cite{5} and characteristics of
some kinds of models are summarized in Alam's works\cite{6}.

The research team of A. G. Riess et al.\cite{7} have pinpointed
the transition epoch from matter domination to dark energy
domination when cosmic expansion began to accelerate and also
indicated the transition between the two epochs is constrained to
be at $z_{T}=0.46\pm 0.13 $. It is therefore pressing that an
independent model of dark energy can be employed to elaborate the
present phenomena and trace back to the past and even forecast the
future evolution. To attain this aim, in this paper we present the
simplest
linear equation of state $p=B(z)\rho $, with $p=p_{dE}$, $%
\rho=\rho_{dm}+\rho_{dE}$, where the pressure of dark matter
$p_{dm}=0$ has been taken into account. This model can predict a
smooth transition from nonrelativistic matter(including both
baryons and dark matter) domination phase to dark energy
domination one in a natural way. Directed by this objective, the
paper is organized as follows. The central part of the paper is
Sec.~2 in which the two constraints in choosing the function $B$
are discussed and a suitable function $B=-\frac{A}{(1+z)^{\alpha
}}$ is proposed. From the function some parameters of the universe
are derived. In Sec.~3 the conclusion is put up that only by
choosing a suitable function $B$ can the unified equation of state
proposed describe the evolution of the universe.

\section{Analysis and deduction}

It is known that our universe begins with the big bang. The
radiation should have been the dominant contribution to the total
density of the universe when the redshift $z$ was more than
$10^{4}$ in the early universe. The density of radiation decreased
and quickly approached $0$ with the evolution of the universe.
However, the density of nonrelativistic matter(including both
baryons and dark matter) increased and became the dominant
contribution to the universe. Subsequently, the density of
nonrelativistic matter decreased and that of dark energy increased
and at the same time cosmic expansion transited from deceleration
to acceleration, the process of which is discussed in this paper.

We assume dark matter and dark energy satisfy the unified equation of state $%
p=B(z)\rho$ with $p=p_{dm}+p_{dE}$, $\rho=\rho_{dm}+\rho_{dE}$, where the
pressure of dark matter and that of dark energy satisfy respectively: $%
p_{dm}=0$, $p_{dE}=\omega _{dE}\rho _{dE}$($\omega _{dE}<0$). Because of the
above equations and the evolution of our universe, there are some
constraints for function $B$ proposed:
\begin{romanlist}[(ii)]
\item Function $B$ should maintain negative value forever and vary
in the range ($\omega_{dE}$, $0$).
\item For $z>0$, function $B$
should have been approaching $0$ but not equaling to $0$ for a
long time in the high redshift and be smaller and smaller as
redshift becomes low.
\end{romanlist}

In terms of the two constraints proposed above, we give a concrete
form of function $B$ as $B=-\frac{A}{(1+z)^{\alpha }} $ and we can
obtain the solutions of some parameters including fractional
components, deceleration-acceleration transition, the equation of
state of dark energy, which are coincident with the evolution of
the universe and consistent with the present astrophysical
observations.

For a spatially flat FRW universe, Friedmann equation can be written as:
\begin{equation}
H^{2}=\frac{8\pi G}{3}\rho _{tot}\,.  \label{equ1}
\end{equation}
Also the continuity equation:
\begin{equation}
\dot{\rho}_{tot}=-3H(p_{tot}+\rho _{tot})\,,  \label{equ2}
\end{equation}
has to be taken into account, where $\rho _{tot}$ is the total
energy density
of our universe $\rho _{tot}=\rho _{b}+\rho _{dm}+\rho _{dE}=\rho _{b}+\rho $%
, where subscript $``b\textquotedblright $ denotes baryons. So
$\rho _{b}$ is the density of baryons and Eq.~(\ref{equ2}) is
separated into two parts:
\begin{equation}
\dot{\rho}_{b}=-3H(p_{b}+\rho _{b})\,,  \label{equ3}
\end{equation}
\begin{equation}
\dot{\rho}=-3H(p+\rho )\,.  \label{equ4}
\end{equation}
If the function $B(z)=-\frac{A}{(1+z)^{\alpha }}$, where $A$,
$\alpha $ are the positive real constants, substituting
$p=B(z)\rho $ to Eq.~(\ref{equ4}) then
\begin{equation}
\dot{\rho}=-3H(-\frac{A}{(1+z)^{\alpha }}\rho +\rho )\,.
\label{equ5}
\end{equation}
So the density of dark matter and dark energy evolves with the
redshift as
\begin{equation}
\rho =C(1+z)^{3}e^{\frac{3A}{\alpha (1+z)^{\alpha }}}\,,
\label{equ6}
\end{equation}
where $C$ is an integral constant. Using the constraint condition
of $z=0$ makes it satisfy:
\begin{equation}
\rho _{0}=Ce^{\frac{3A}{\alpha }}\,,  \label{equ7}
\end{equation}
where $\rho _{0}$ is the present density of dark matter and dark energy and
today's evaluated quantities will hereafter denoted by the label $%
``0\textquotedblright $, and
\begin{equation}
\rho _{c0}=\rho _{tot0}=\rho _{0}+\rho
_{b0}=\frac{3H_{0}^{2}}{8\pi G}\,, \label{equ8}
\end{equation}
is how the present critical energy density is defined. Replacing Eq.~(\ref%
{equ8}) in Eq.~(\ref{equ7}), we obtain:
\begin{equation}
C=(\rho _{c0}-\rho _{b0})e^{-\frac{3A}{\alpha }}\,.  \label{equ9}
\end{equation}
The density of baryons and that of dark matter are respectively recognized
as:
\begin{equation}
\rho _{b}=\rho _{b0}(1+z)^{3}\,,  \label{equ10}
\end{equation}
\begin{equation}
\rho _{dm}=\rho _{dm0}(1+z)^{3}\,.  \label{equ11}
\end{equation}
So the total density takes the form:
\begin{equation}
\rho _{tot}=(\rho _{c0}-\rho _{b0})(1+z)^{3}e^{\frac{3A}{\alpha
(1+z)^{\alpha }}-\frac{3A}{\alpha }}+\rho _{b0}(1+z)^{3}\,,
\label{equ12}
\end{equation}
and the density of dark energy can be obtained by
Eqs.~(\ref{equ10}), (\ref {equ11}) and (\ref{equ12}) as:
\begin{equation}
\rho _{dE}=(\rho _{c0}-\rho _{b0})(1+z)^{3}e^{\frac{3A}{\alpha
(1+z)^{\alpha }}-\frac{3A}{\alpha }}-\rho _{dm0}(1+z)^{3}\,.
\label{equ13}
\end{equation}
One can express the fractional energy densities $\Omega _{b}$,
$\Omega _{dm}$ , $\Omega _{dE}$ as
\begin{equation}
\ \Omega _{b}=\frac{\ \Omega _{b0}}{(1-\Omega
_{b0})e^{\frac{3A}{\alpha (1+z)^{\alpha }}-\frac{3A}{\alpha
}}+\Omega _{b0}}\,,  \label{equ14}
\end{equation}
\begin{equation}
\ \Omega _{dm}=\frac{\ \Omega _{dm0}}{(1-\Omega
_{b0})e^{\frac{3A}{\alpha (1+z)^{\alpha }}-\frac{3A}{\alpha
}}+\Omega _{b0}}\,,  \label{equ15}
\end{equation}
\begin{equation}
\ \Omega _{dE}=\frac{\ (1-\Omega _{b0})e^{\frac{3A}{\alpha
(1+z)^{\alpha }}- \frac{3A}{\alpha }}-\Omega _{dm0}}{(1-\Omega
_{b0})e^{\frac{3A}{\alpha (1+z)^{\alpha }}-\frac{3A}{\alpha
}}+\Omega _{b0}}\,,  \label{equ16}
\end{equation}
where $\Omega _{b0}=\frac{\rho _{b0}}{\rho _{c0}}$, $\Omega
_{dm0}=\frac{ \rho _{dm0}}{\rho _{c0}}$.

Assuming $A=0.7$ and $\alpha =1.81$, the fractional energy
densities are only the functions of redshift with the prior of
$\Omega _{b0}=0.04 $, $ \Omega _{dm0}=0.3 $. The assumption is
still used in the following calculation.

\begin{figure}[th]
\centerline{\psfig{file=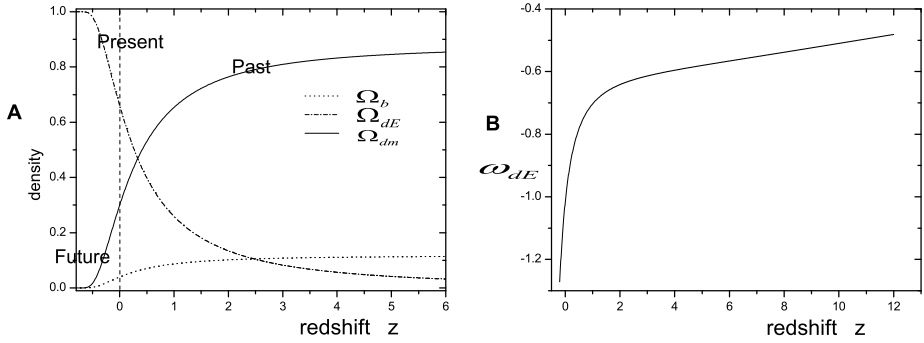,width=5.5in}} \vspace*{8pt}
\caption{A: The fractional energy densities as the functions of
redshift. B: The equation of state of dark energy $\protect\omega
_{dE}$ as the function of redshift. The priors $\Omega
_{b0}=0.04$, $\Omega _{dm0}=0.3$ have been used.} \label{fig1}
\end{figure}

The results of the three fractional energy densities at the
present time are compatible with observations using the priors of
$\Omega _{b0}=0.04$, $\Omega _{dm0}=0.3$. Meanwhile, according to
FIG.~1A we know that the past evolution of baryons, dark matter
and dark energy is consistent with what is recognized. The
cosmology will end up with a Big Rip in that the universe will be
filled with dark energy in future. The conclusion is shared in
many literatures(e.g.\cite{8}).

It follows that the equation of state parameter of dark energy can be
obtained as:
\begin{equation}
\omega _{dE}=\frac{p_{dE}}{\rho _{dE}}=-\frac{A}{(1+z)^{\alpha }}\frac{%
(1-\Omega _{b0})e^{\frac{3A}{\alpha (1+z)^{\alpha }}-\frac{3A}{\alpha }}}{%
(1-\Omega _{b0})e^{\frac{3A}{\alpha (1+z)^{\alpha
}}-\frac{3A}{\alpha }}-\ \Omega _{dm0}}\,.  \label{equ17}
\end{equation}
Using Eq. (\ref{equ17}) we can plot $\omega _{dE}$ as the function
of redshift(see FIG.~1B). The equation of state of dark energy is
$\omega _{dE0}\approx-1.019$ at present. Tracing back to the past,
the larger the redshift is, the more the density of
nonrelativistic matter(including both baryons and dark matter) is
and the more adjacent to $0$ the equation of state of dark energy
$\omega _{dE}$ is, i.e., we realize an evolving dark energy with
the equation of state being below $-1$ around the present epoch
evolved from $\omega _{dE}>-1$ in the past\cite{9}. This is
another consistency between our theory and the evolution.

The dominance of the dark energy leads to the acceleration
expansion of our universe. The increasing density of dark energy
is the reason why expansion of our universe transited from
deceleration to acceleration. Our theory describes the transition.
The deduction is as follows:
\begin{equation}
q=-\frac{\ddot{a}a}{\dot{a}^{2}}=-\frac{\ddot{a}}{aH^{2}}\,,
\label{equ18}
\end{equation}%
where $q$ is the deceleration parameter. Using Eqs.~(\ref{equ1}), (\ref{equ6}%
) and (\ref{equ9}), the Friedmann equation may be rewritten as
\begin{equation}
H^{2}=\frac{8\pi G}{3}(\rho _{c0}-\rho
_{b0})(1+z)^{3}e^{\frac{3A}{\alpha (1+z)^{\alpha
}}-\frac{3A}{\alpha }}+\frac{8\pi G}{3}\rho _{b0}(1+z)^{3}\,.
\label{equ19}
\end{equation}%
Differentiating Eq.~(\ref{equ19}) with respect to time and
changing variable from the cosmic time to the redshift, we finally
obtain:
\begin{equation}
\frac{\ddot{a}}{a}=-\frac{1}{2}H^{2}+\frac{3A(\rho _{c0}-\rho _{b0})}{%
2(1+z)^{\alpha -3}}e^{\frac{3A}{\alpha (1+z)^{\alpha
}}-\frac{3A}{\alpha }}\,. \label{equ20}
\end{equation}%
Combining the above two equations, we find that $q$ satisfies
\begin{equation}
q=\frac{1}{2}-\frac{3A}{2(1+z)^{\alpha }}\frac{\ (1-\Omega _{b0})e^{\frac{3A%
}{\alpha (1+z)^{\alpha }}-\frac{3A}{\alpha }}}{(1-\Omega _{b0})e^{\frac{3A}{%
\alpha (1+z)^{\alpha }}-\frac{3A}{\alpha }}+\Omega _{b0}}\,.
\label{equ21}
\end{equation}%
From this expression, we get the present deceleration parameter $%
q_{0}=-0.508$ according to $z=0$ and the transition redshift $z_{T}\approx
0.449$ according to $q=0$. The deceleration parameter as the function of
redshift is shown in FIG.2.
\begin{figure}[th]
\centerline{\psfig{file=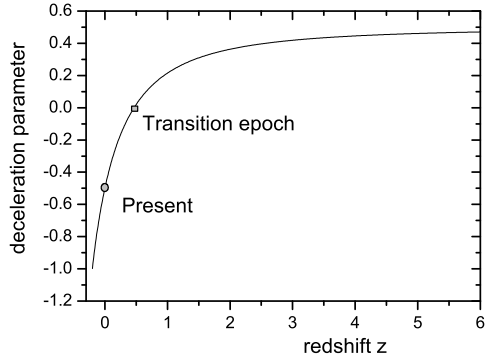,width=3.0in}} \vspace*{8pt}
\caption{The deceleration parameter $q$ as the function of redshift. The
priors $\Omega _{b0}=0.04$, $\Omega _{dm0}=0.3$ have been used.}
\label{fig2}
\end{figure}

From above deduction, we know the acceleration of our cosmic
expansion will be faster and faster when
$B=-\frac{A}{(1+z)^{\alpha }}$ is chosen. As a result of smaller
and smaller $B$, $q$ and $\omega_{dE}$ will also be smaller and
smaller and even will approach negative infinity from FIG.~3. It
is inevitable that the universe will end up with a Big Rip. On the
condition that choosing another function based on the above two
constraints and using another symbol: $B_{2}=-0.7exp(-z^{2})$ and
($z>-1$) in order to distinguish the two functions, we can still
obtain the solutions coincident with the past and the present
observational data because of the semblable curves of the two
functions in $z>0$. But the fate of the universe will be
different. New function $B_{2}$ is axial symmetry. Smaller $z$
will lead to larger $B_{2}$ in future which also reflects the
tendency of the equation of state of dark energy. Both of the
densities of baryons, dark matter will approach small quantities,
so we think that the total energy density approximately equal to
the density of dark energy. Because of
\begin{equation}
q=-\frac{\ddot{a}a}{\dot{a}^{2}}=-\frac{\ddot{a}}{aH^{2}}=\frac{1}{2}+\frac{3%
}{2}\frac{p_{tot}}{\rho _{tot}}\,,
\end{equation}%
one obtains
\begin{equation}
q\approx\frac{1}{2}+\frac{3}{2}B_{2}\,.
\end{equation}%
When $B_{2}>-\frac{1}{3}$, $q>0$, the expansion of the universe
will come back to deceleration from acceleration. We don't discuss
the detailed deduction owing to the difficulty of the integral of
Gaussian function.
\begin{figure}[th]
\centerline{\psfig{file=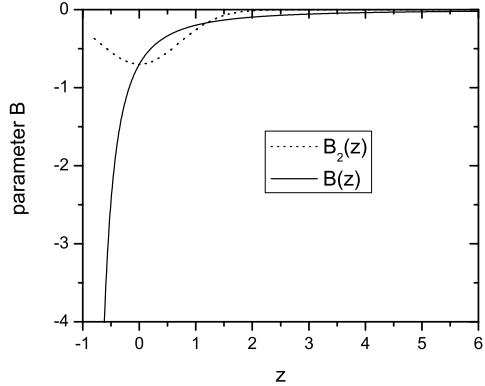,width=3.0in}} \vspace*{8pt}
\caption{The contrast of two kinds of the choices of function $B$}
\label{fig3}
\end{figure}

As shown in the above discussion, the results can be obtained which are in
agreement with the observational data of the past time and the present time
as long as we choose the function $B$ in terms of the above two analyzed
constraints. At the same time the tendency of B in the range $-1<z<0$
determines the fate of the universe. So it is possible to find a suitable
function to describe the evolution of the universe from the early time to
the future.

\section{conclusion}

In summary, this paper considers that the utilization of a unified
linear equation of state can describe the history of the universe
and its evolution of the future. It is worth highlighting that we
don't simulate function $B$ without foundation but based on some
remarkable constraints. Analyzing the constraints of the choice of
the function $B$ in terms of the evolution of our universe and
choosing the appropriate function $B$ according to the
constraints, the evolution of our universe can be obtained in
consistence with what is recognized. Both dark matter and dark
energy are considered the essential but missing pieces in the
cosmic jiasaw puzzle\cite{10}. But the nature of either dark
matter or dark energy is currently unknown. We propose this
model-independent unified equation of state which can be utilized
to forecast the nature of dark matter and dark energy possibly.

\section*{Acknowledgments}

We are grateful to Prof. Hongya Liu, Dr. Lixin Xu and Mr. Hongjun
Chen for helpful discussions. This work is supported by National
Natural Science Foundation of China under Grant NO.10275008.


\begin{thebibliography}{99}
\bibitem{1} C.~L.~Bennett \textit{et al.}, {\it Astrophys. J. Suppl. Ser.} \textbf{%
148}, 1 (2003)

\bibitem{2} K. Freese \textit{et al.}, {\it Nucl.Phys.} \textbf{B287}, 797 (1987);
P. J. E Peebles and B. Ratra, {\it ApJ} \textbf{325}, L17 (1988)

\bibitem{3} Pedro F. Gonzalez-Diaz and Carmen L. Siguenza, {\it
Nucl.Phys.}
\textbf{B697}, 363 (2004)

\bibitem{4} L. Mersini, M. Bastero-Gil, and P. Kanti, {\it PRD} \textbf{64},
043508 (2001); K. Freese and M. Lewis, {\it Phys.Lett.B}
\textbf{540}, 1 (2002)

\bibitem{5} M. C. Bento and O. Bertolami, {\it PRD} \textbf{70},
083519 (2004)

\bibitem{6} Ujjaini Alam \textit{et al.}, {\it Mon.Not.Roy.Astron.Soc} \textbf{347},
L47 (2004)

\bibitem{7} A. G. Riess \textit{et al.}, {\it Astrophys.J.} \textbf{607},
665 (2004)

\bibitem{8} Shin'ichi Nojiri, Sergei D. Odintsov, {\it Phys.Lett.} \textbf{B595},
1 (2004)

\bibitem{9} Bo Feng, Xiulian Wang, and Xinmin Zhang, astro-ph/0404224

\bibitem{10} Varun Sahni, astro-ph/0403324
\end{thebibliography}
\end{document}